# Recent increase in near-perpendicular tropical cyclone landfall fraction projected to emerge under high-emissions warming

Cong Gao[1*], Ning Lin[1*]

**Affiliations:**

[1]Department of Civil and Environmental Engineering, Princeton University, Princeton, NJ, USA

[*]Corresponding authors. Email: cong.gao@princeton.edu (C.G.); nlin@princeton.edu (N.L.)

**Tropical cyclone landfall angle strongly modulates storm-surge potential, but how it changes with climate remains unknown. From 1980 to 2024, the global fraction of near-perpendicular landfalls shows a weak, statistically insignificant increase, with no significant basin-scale trends. PepC-Global simulations driven by reanalyses reproduce observed landfall-angle and pre-landfall approach-angle distributions and broad geographic variations in the near-perpendicular fraction. When driven by output from 14 CMIP6 models, the simulations show a similarly weak and statistically insignificant increase during 1980–2024, suggesting a forced contribution not yet detectable above internal variability. Under high emissions, however, the global near-perpendicular fraction increases significantly through 2099, with the strongest and most spatially coherent changes along the eastern North Pacific, whereas no significant global trend emerges under intermediate emissions. The projected increase arises mainly from changes in near-perpendicular fractions at individual coastal locations rather than from a redistribution of landfalls. These results identify landfall angle as an overlooked dimension of changing tropical cyclone coastal risk, one whose evolution this century depends on the emissions pathway.**

## INTRODUCTION

Tropical cyclones (TCs) are among the most damaging natural hazards globally, particularly in coastal regions that are often densely populated and highly developed[1–4]. The greatest impacts from these storms typically occur at or near the time of landfall, when multiple hazards converge, with storm surge representing one of the most destructive and deadly components[5,6]. The landfall angle, defined here as the angle between the TC track at landfall and the local coastline orientation (Extended Data Fig. 1), plays a critical role in determining the magnitude of storm surge[7–10], as approaches with a stronger cross-shore component drive water onshore far more effectively than those dominated by alongshore motion. For example, Hurricane Katrina made landfall in 2005 at a close to perpendicular angle and produced the highest storm surge recorded globally by TCs since 1980, with a maximum high-water mark of 8.5 m (27.9 ft) at Pass Christian, Mississippi[11]; as one of the costliest hurricanes in United States history, it caused more than USD 100 billion in losses[1] and resulted in approximately 2,000 fatalities[12]. Hurricane Sandy (2012) reinforces the point: its track crossed the New Jersey coastline at an angle closer to perpendicular than any previous hurricane on record, an unusual approach that drove record-setting peak water levels across New Jersey and New York[13], inundated lower Manhattan and the New York City subway system, and caused over USD 70 billion in damages and more than 200 fatalities[14,15]. Although previous studies have investigated the frequency of Hurricane Sandy-like landfalls in the New Jersey–New York region[13,16], whether near-perpendicular landfalls have become more frequent globally or in individual basins in recent decades, and how they will evolve under continued anthropogenic warming, remain unknown.

Observed trends[17–21] and future projections[22,23] of TC activity have been the subject of a large and growing body of research. The proportion of storm fixes reaching major-hurricane intensity has increased[24], rapid intensification has become more common[25], and TC rainfall rates have risen[26], consistent with thermodynamic expectations. Projections of global TC genesis frequency have also attracted considerable attention[22,23,27,28], yet remain uncertain, with most climate models suggesting a modest decrease under continued warming. Comparatively less attention has been paid to changes in TC tracks, and most of this work has focused on track occurrence and translation speed. Track-related changes have nevertheless emerged, including a poleward migration of TC lifetime maximum intensity[29], a recent migration of TCs towards coasts[30] and a reported slowdown of TC translation speed[31], although the latter remains under debate[32,33]. Detecting signals in TC landfall is even more difficult because landfalls are relatively rare[21,23,34]. Moreover, landfall angle has been examined mainly through storm-surge sensitivity studies[7–10] and individual-event analyses, most notably Hurricane Sandy[13], leaving its global variability, trends and future response largely unknown.

We focus on five major continental coastline regions: the North Indian Ocean, western North Pacific, eastern North Pacific, North Atlantic and Australian Region. These regions together cover the main open-ocean TC-prone continental coasts. Island coastlines are excluded because their small size, complex geometry and sensitivity to coastline resolution can introduce large uncertainty in local coastline orientation. The angle concept is also less physically meaningful there: narrow shelves and deep adjacent water limit wind-driven surge, so total coastal water level is instead governed largely by wave setup and runup, which depend more on how far the storm passes from the island and which shore faces its strongest winds than on the angle between the track and the coast. African coastlines are not included because the TC-prone African coast mainly faces the semi-enclosed Mozambique Channel rather than the open ocean.

Whether the global frequency of near-perpendicular landfalls carries a detectable forced signal, and whether one emerges under continued warming, has not been established. Here we assess recent and projected changes in the fraction of near-perpendicular TC landfalls across these five regions. We combine observed landfalls from 1980 to 2024 with simulations from the Princeton Environment-Dependent Probabilistic Tropical Cyclone Global model (PepC-Global)[35], driven by reanalyses and by output from 14 CMIP6 models under two emissions scenarios. We find no detectable global increase over 1980–2024, but a significant increase emerging by 2099 under high-emissions warming.

## RESULTS

### Observed weak increase in near-perpendicular landfall fraction

We first examine observed changes in the fraction of near-perpendicular TC landfalls from 1980 to 2024. Using IBTrACS[36], we identify 1,036 TCs that made landfall along these five continental coastlines over this period. Landfall angles are calculated as described in Methods and illustrated in Extended Data Figs. 1 and 2; following Hall and Sobel[13], angles of at least 60° are classified as near-perpendicular. We define the near-perpendicular fraction as the number of near-perpendicular landfalls divided by the total number of landfalls. Computed in 3-year bins, the global near-perpendicular fraction shows a weak, statistically insignificant increase of 0.050 percentage points per year ($p = 0.671$; Fig. 1g). The global fraction fluctuates markedly among the 3-year bins, indicating that the fitted increase is small relative to the large observed temporal variability. Local trends, computed from the near-perpendicular fraction within each along-coast analysis window and 3-year bin, are spatially heterogeneous, with distinct clusters of increases and decreases (Fig. 1a). These spatial contrasts indicate that the weak global increase reflects compensation among opposing local tendencies rather than a coherent increase across the analysed coastlines. Consistent with these local patterns, the basin-scale trend is positive in the North Indian Ocean (0.309 percentage points per year; $p = 0.426$; Fig. 1b), negative in the western North Pacific (−0.100 percentage points per year; $p = 0.528$; Fig. 1c), negative in the eastern North Pacific (−0.330 percentage points per year; $p = 0.327$; Fig. 1d), positive in the North Atlantic (0.469 percentage points per year; $p = 0.112$; Fig. 1e) and positive in the Australian Region (0.030 percentage points per year; $p = 0.922$; Fig. 1f). None of the global or basin-scale trends is statistically significant, indicating that the observed changes are weak and spatially mixed rather than a robust recent shift in near-perpendicular landfall fraction.

**Validation of PepC-Global on landfall angles**

We use PepC-Global[35] (see Methods), a probabilistic, physics-grounded model that efficiently generates large ensembles of synthetic TCs. The probabilistic formulation permits the generation of millions of plausible TC tracks given the large-scale environments, providing a sufficiently large sample of landfalls to characterize their angle distributions robustly. PepC-Global has been shown to reproduce observed landfall frequency and landfall intensity well[35], but its ability to reproduce observed landfall angles has not been evaluated; to the best of our knowledge, the same is true for other models of this type, including the Columbia HAZard (CHAZ) model[37] and the statistical-deterministic model[38,39]. Across 400 stochastic realizations of PepC-Global driven by atmospheric reanalysis ERA5[40] and ocean reanalysis GODAS[41] (see Methods) over 1980–2024, the model reproduces the observed probability density of sin(landfall angle). We characterize landfall angle by its sine because it directly represents the normalized cross-shore component of the approach direction—ranging from 0 for coast-parallel to 1 for perpendicular

landfall—and is therefore more physically relevant to storm-surge potential. The observed global distribution lies within the 95% simulation envelope across the full range, including the near-perpendicular tail that defines this study's central metric (Fig. 2a). Similar agreement is found along each of the five continental coastlines (Extended Data Fig. 3). Note that near-perpendicular landfalls are common among landfalling TCs at both global and basin scales, indicating that changes in their relative frequency are a globally relevant question for coastal risk. PepC-Global also reproduces spatial variations in the climatological near-perpendicular landfall fraction. Simulated and observed local fractions, evaluated at the same spatial analysis windows as in Fig. 1a, are significantly correlated ($R = 0.47$, $p = 4.91 \times 10^{-5}$; Fig. 2b). The points broadly follow the one-to-one line, indicating that PepC-Global generally captures the contrast between coastal locations with lower and higher observed fractions, although local deviations remain. The corresponding maps further show similar large-scale spatial patterns (Fig. 2c,d). Because PepC-Global generates TC tracks from large-scale environmental conditions, this agreement also indicates that large-scale environmental controls on TCs explain much of the observed spatial variation in landfall angles. PepC-Global is therefore well suited to characterizing landfall-angle behaviour across scales, from global to local, and to examining how changes in the large-scale environment under warming alter landfall angles. The PepC-Global simulation is also useful for risk analysis, particularly for regions with few landfalling TCs.

We also test whether PepC-Global can reproduce the net approach angles shortly before landfall (see Methods). At global scale, the observed probability densities of sin(6-h net pre-landfall approach angle) and sin(12-h net pre-landfall approach angle) lie within the 95% simulation envelope of the 400 PepC-Global realizations, including the near-perpendicular tail (Fig. 3a,b). Similar agreement is found along each of the five continental coastlines at both 6 h (Extended Data Fig. 4) and 12 h (Extended Data Fig. 5) before landfall. Because landfall can occur at any time between 6-hourly observations, the landfall angle represents storm motion during the 0–6 h preceding landfall relative to the coastline segment crossed. Agreement for the 6- and 12-h net approach angles shows that PepC-Global also captures the incoming trajectory over longer pre-landfall intervals. This consistency strengthens confidence in using PepC-Global to assess changes in landfall angles.

**Possible forced contribution during 1980–2024**

We next assess whether the weak observed changes are consistent with an externally forced response. PepC-Global is driven by output from 14 CMIP6 models (Extended Data Table 1), combining historical simulations for 1980–2014 with simulations for 2015–2024 under SSP245, an intermediate-emissions scenario of the Shared Socioeconomic Pathways (SSP2 with a radiative forcing of 4.5 W $m^{-2}$ by 2100). Pooling landfalls across the 14 models and 100 stochastic realizations per model suppresses internal and sampling variability, emphasizing the common forced response. The pooled global fraction, computed in 3-year bins, shows a weak, statistically insignificant increase in the near-perpendicular landfall fraction (slope = 0.013 percentage points per year, $p = 0.222$; Fig. 4a), consistent in sign with the observed increase (Fig. 1g). Most individual models (11 of 14) also show positive global trends, although these are masked by large variability across stochastic realizations (Fig. 4b). Local trends are spatially mixed, and no analysis centre remains significant after false-discovery-rate (FDR) control (Fig. 4c). The consistent sign of the observed and simulated global trends and the predominance of positive trends across models suggest that external forcing may already contribute weakly to the observed changes, although this contribution is not yet detectable above internal variability.

**Forced increase emerges under high emissions**

We next examine whether a forced response in near-perpendicular landfall fraction emerges under continued warming, particularly under a high-emissions scenario. PepC-Global is driven by output from the same 14 CMIP6 models, with 100 stochastic realizations generated for each model and landfalls pooled across all models and realizations in 3-year bins. Under SSP585, a high-emissions scenario (SSP5 with a radiative forcing of 8.5 W $m^{-2}$ by 2100), the pooled global fraction increases significantly during 2015–2099 (slope = 0.015 percentage points per year, $p < 0.001$; Fig. 5a). Local trends are strongest and most spatially coherent along the eastern North Pacific, where 13 of 16 analysis centres show increases, including the only two of all 80 analysis centres that remain significant after FDR control, and the three negative trends occur at low latitudes, whereas changes elsewhere are smaller and spatially mixed (Fig. 5b). Along the northern coast of Australia, local trends diverge, with increases in the west and decreases in the east (Fig. 5b). Increases are also projected along the Yellow Sea–Bohai coast of northern China and the US Mid-Atlantic coast, implying that, for a given landfall frequency, near-perpendicular landfalls such as Hurricane Sandy (2012) will become more likely, a risk that is easily underestimated because TC landfalls in these regions are rarer than along lower-latitude coastlines (Fig. 5b).

The response is substantially weaker under SSP245. The global trend is small and statistically insignificant (slope = 0.002 percentage points per year, $p = 0.651$; Fig. 5c). Locally, the Bay of Bengal shows spatially coherent increases in the near-perpendicular landfall fraction (Fig. 5d). The global SSP585–SSP245 difference increases significantly over time (slope = 0.014 percentage points per year, $p = 0.014$; Fig. 5e). Most coastal centres along the eastern North Pacific and North Atlantic show positive local trend differences, highlighting a larger increase in the near-perpendicular landfall fraction along North American coastlines under high emissions than under intermediate emissions (Fig. 5f). These results indicate that a detectable forced increase in near-perpendicular landfalls emerges under high emissions but not under the intermediate-emissions pathway. Together with the absence of a detectable trend over 1980–2024 (Fig. 4a), the stronger response under SSP585 indicates that detection of the forced signal depends on the magnitude of warming. Sustained high emissions may therefore increase near-perpendicular landfalls and associated coastal hazards, particularly storm surge.

## DISCUSSION

Our results show that the increase in the fraction of near-perpendicular TC landfalls becomes larger and more detectable as climate forcing increases. During 1980–2024, observations exhibit a weak, statistically insignificant global increase in the near-perpendicular landfall fraction, accompanied by large temporal variability, spatially mixed local trends and no significant basin-scale changes (Fig. 1). PepC-Global reproduces the observed landfall-angle and pre-landfall approach-angle distributions, as well as broad geographic variations in the near-perpendicular landfall fraction, supporting its use for climate projections (Figs. 2 and 3). CMIP6-driven simulations show a similarly weak and statistically insignificant global increase over the same period, with positive trends in 11 of 14 models, suggesting a weak but not yet detectable forced contribution (Fig. 4). Under SSP585, however, a significant global increase emerges through 2099, with the strongest and most spatially coherent changes along the eastern North Pacific; under SSP245, the global trend remains insignificant and the local increases are generally weaker and less extensive (Fig. 5). To our knowledge, this is the first global-scale assessment of observed and projected changes in TC landfall angle, a key geometric influence on storm-surge potential.

Explaining the projected increase is challenging because landfall angle is the end product of a chain of factors, including the location and timing of TC genesis, the steering flow that governs translation direction, the rate at which storms intensify and decay along their tracks, and the orientation of the coastline they ultimately encounter. A change in any one of these factors, or in their interaction, could plausibly shift the distribution of landfall angles, and these factors are themselves coupled in ways that are not yet well understood under sustained warming. The projected global increase can arise in two ways: the near-perpendicular fraction can change at individual centres (Fig. 5), or landfalls can shift towards centres where near-perpendicular approaches are already more common. Fig. 6a,b shows a redistribution of landfalls among centres under both scenarios, but this redistribution does not preferentially shift landfalls towards centres with historically high near-perpendicular fractions. Across the 80 analysis centres, changes in each centre's share of global landfalls are weakly and insignificantly correlated with its simulated historical (1980–2009) near-perpendicular fraction under SSP245 ($R = 0.11$, $p = 0.339$; Fig. 6c). Under SSP585, the correlation is similarly weak and insignificant ($R = 0.14$, $p = 0.227$; Fig. 6d). Overall, although landfalls are redistributed among centres, this redistribution does not systematically favour centres with high near-perpendicular fractions, suggesting that the projected global increase arises primarily from changes in near-perpendicular fractions at individual centres rather than from shifts in where landfalls occur.

These projections should be read in the context of scenario plausibility. SSP585 is a high-end forcing pathway that recent assessments regard as unlikely to be realized, and it is better understood as an upper bound on twenty-first-century warming than as a most-likely trajectory[42]. The results remain informative precisely because the response is scenario-dependent: a detectable global increase in near-perpendicular landfalls emerges under SSP585 but not under SSP245 over the same period, indicating that whether this component of coastal TC risk changes materially this century depends on the emissions pathway followed. Emissions consistent with an intermediate rather than a high-end pathway would therefore be expected to limit the projected increase.

Landfall angle is an underexamined but potentially important dimension of TC risk in a changing climate, with implications that extend from long-term coastal adaptation to operational prediction. A key next step is to test whether the projected increase in near-perpendicular landfalls is reproduced across a wider range of modelling frameworks, including high-resolution climate models[43], climate-conditioned AI weather models such as WeatherNext Cyclone[44], emerging AI climate models[45] and other statistical or statistical–dynamical downscaling methods[37–39]. Such projections should first be evaluated against observed landfall angles. Future work should also quantify how the projected increase in near-perpendicular landfalls translates into coastal impacts, particularly extreme storm surge. Finally, although our analysis focuses on long-term climate change, landfall angle may also provide useful information at seasonal-prediction and weather-forecast timescales. Explicitly evaluating landfall-angle prediction skill, and improving it where possible, could therefore strengthen both climate-scale hazard assessment and near-term coastal risk preparedness.

## MAIN REFERENCES

## FIGURES

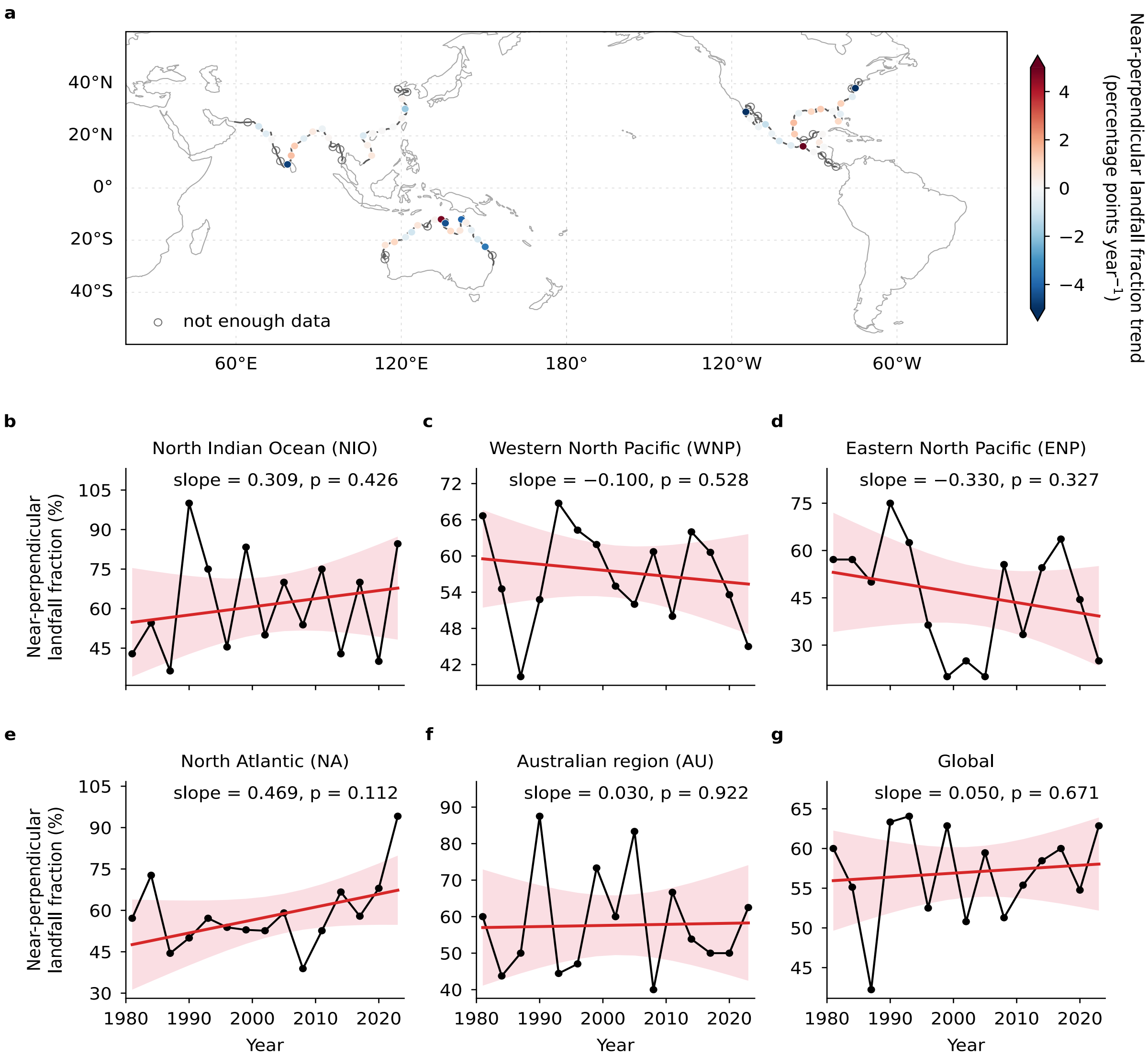

**Fig. 1 | Observed trends in the fraction of near-perpendicular tropical cyclone landfalls. a**, Local trends along the analysed coastlines. **b**, North Indian Ocean (NIO). **c**, Western North Pacific (WNP). **d**, Eastern North Pacific (ENP). **e**, North Atlantic (NA). **f**, Australian Region (AU). **g**, Global. In **a**, coloured circles show the local trend in the fraction of landfalls with near-perpendicular angles (see Methods) at analysis centres spaced every 500 km along the coastlines (thick black lines). Each circle marks the centre of an along-coast window. The window extends 250 km along the coastline on each side of the centre, for a total arc length of 500 km. All window distances are measured along the coastline rather than as straight-line distance from the centre. Unfilled grey circles mark windows with fewer than three 3-year bins containing landfalls, for which no trend is estimated. Of the 80 analysis centres, 57 have estimated trends (coloured circles) and 23 have no trend (unfilled grey circles). Marker colour gives the weighted least-squares slope in percentage points per year, with blue indicating a decreasing and red an increasing fraction. In

**b**–**g**, time series show the near-perpendicular landfall fraction. Each point represents a 3-year bin over 1980–2024 for a total of 15 points. Black lines connect the observed 3-year fractions; red lines show weighted least-squares trends with associated slopes and p-values displayed in each panel; pink shading indicates the 95% confidence interval around the trend.

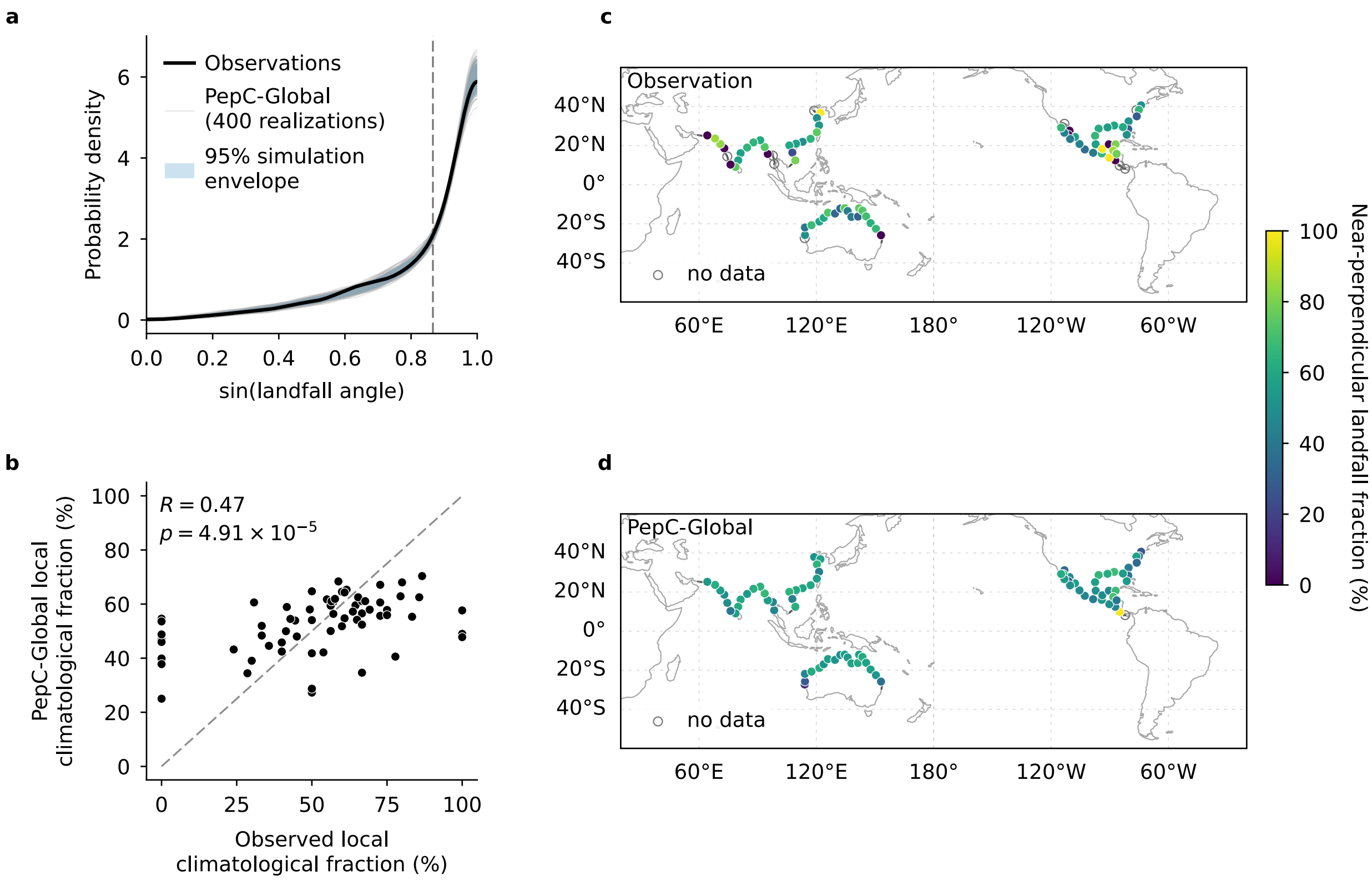

**Fig. 2 | Observed and PepC-Global simulated tropical cyclone landfall angle distributions and near-perpendicular landfall fractions. a**, Probability density of sin(landfall angle). **b**, Observed versus simulated local climatological fraction of landfalls with near-perpendicular angles (see Methods). **c**, Observations. **d**, PepC-Global. In **a**, the black line shows observations, each grey line one of the 400 realizations from PepC-Global driven by ERA5 and GODAS, and the shaded band the 95% simulation envelope; the dashed line marks the 60° near-perpendicular threshold. In **b**–**d**, climatological near-perpendicular landfall fractions are evaluated at the same spatial analysis windows as in Fig. 1a; **b** compares them at each analysis centre (Pearson correlation and p shown); **c** and **d** map them on a common colour scale.

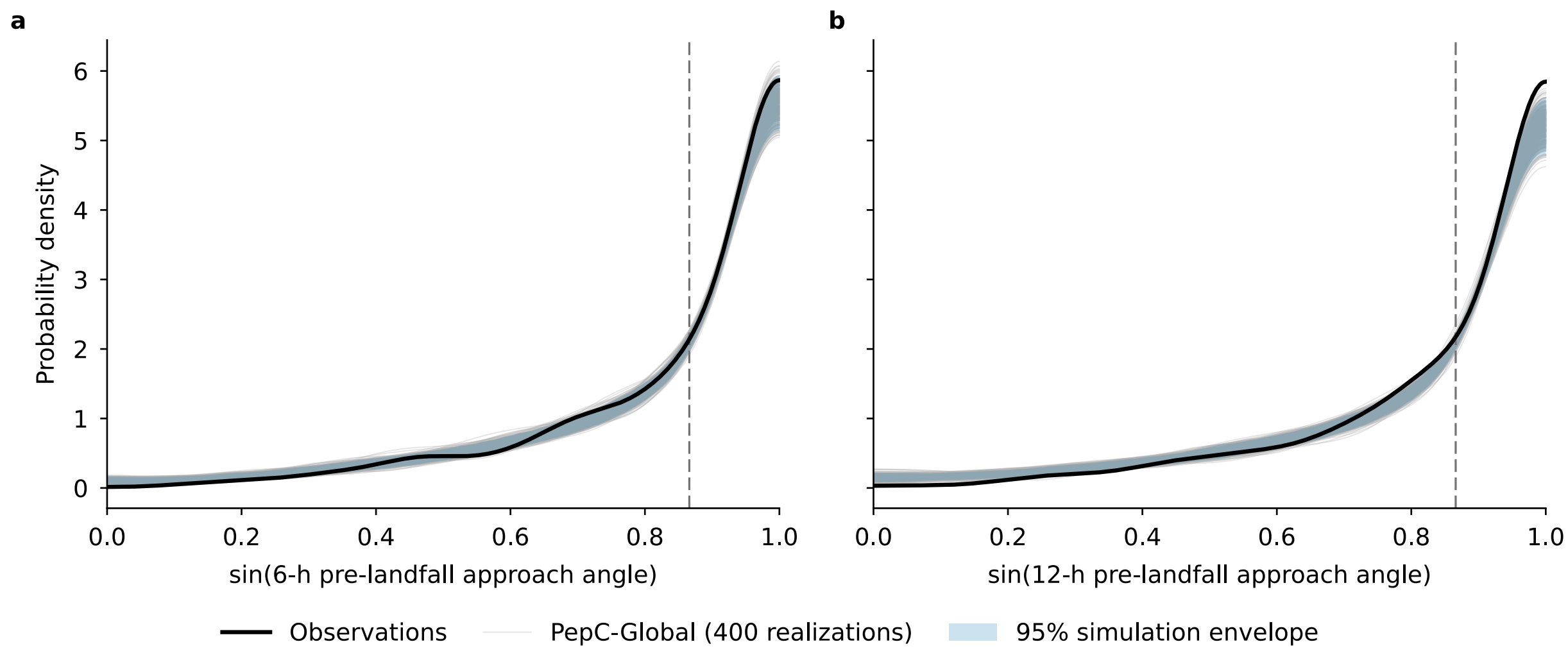

**Fig. 3 | Observed and PepC-Global simulated distributions of tropical cyclone net pre-landfall approach angles. a**, Probability density of sin(6-h net pre-landfall approach angle). **b**, Probability density of sin(12-h net pre-landfall approach angle). See Methods for definitions of the net pre-landfall approach angles. In both panels, the black line shows observations, each grey line one of the 400 realizations from PepC-Global driven by ERA5 and GODAS, and the shaded band the 95% simulation envelope; the dashed line marks the 60° near-perpendicular threshold.

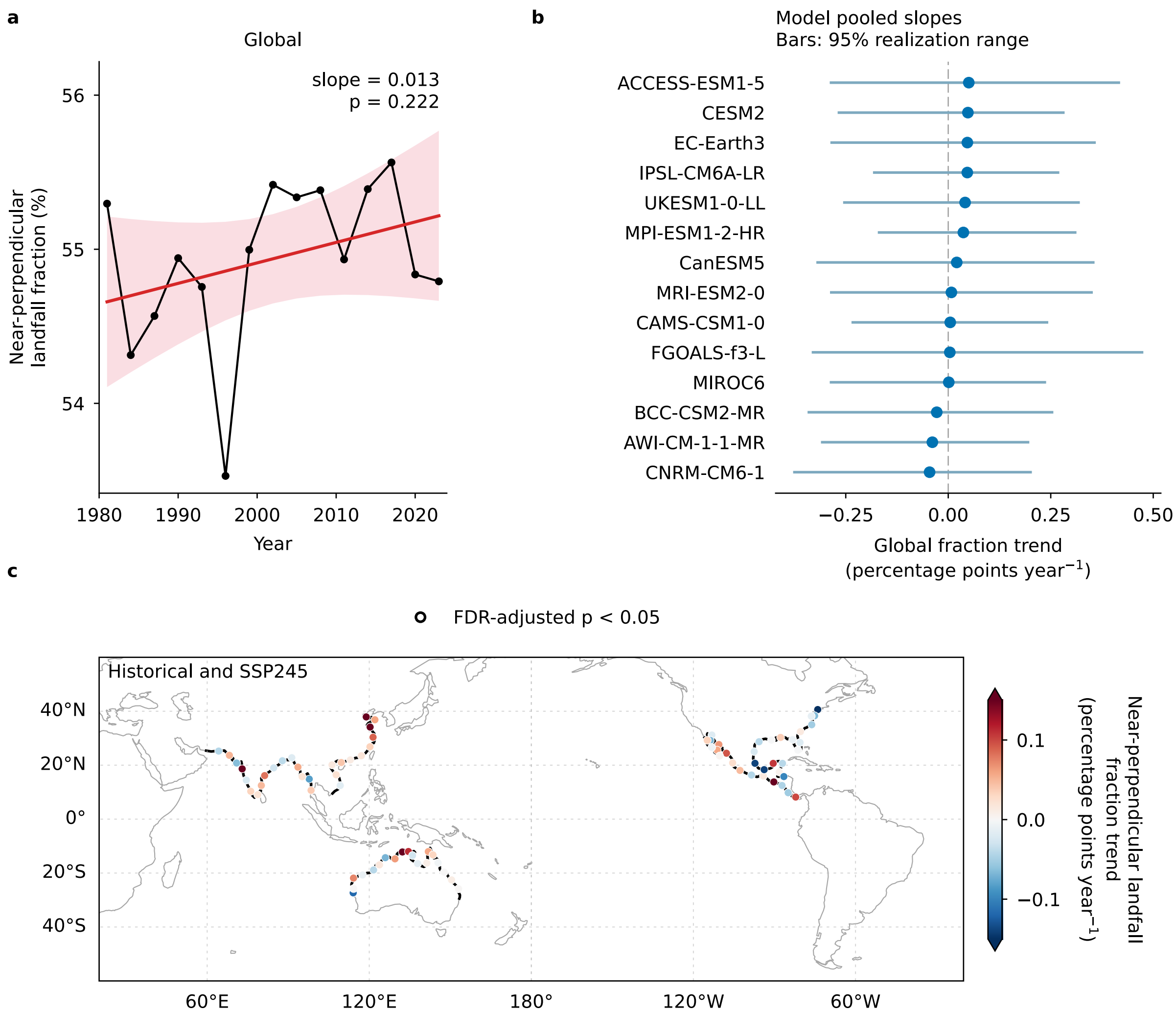

**Fig. 4 | Simulated changes in the fraction of near-perpendicular tropical cyclone landfalls during 1980–2024. a**, Global near-perpendicular landfall (see Methods) fraction in pooled 3-year bins, combining historical simulations for 1980–2014 with SSP245 simulations for 2015–2024. Each point represents a 3-year bin pooled across 100 PepC-Global realizations for each of the 14 CMIP6 models (Extended Data Table 1). Black lines connect the pooled 3-year fractions; the red line shows the weighted least-squares trend, and pink shading indicates its 95% confidence interval. **b,** Model-specific global trends over 1980–2024. Blue circles show slopes calculated from landfall fractions pooled across the 100 PepC-Global realizations for each CMIP6 model. Horizontal bars show the 2.5th–97.5th percentiles of slopes across the 100 individual realizations, and the dashed vertical line marks zero trend. Slopes are expressed in percentage points per year. **c,** Local trends over the combined historical–SSP245 period. At each centre, landfall counts are pooled across all 14 CMIP6 models and their 100 realizations, and a single trend is fitted to the pooled fractions. Each marker shows this pooled trend in percentage points per year, with blue indicating a decrease and red an increase in the fraction. Black outlines mark centres where the trend remains significant at $p < 0.05$ after controlling the false discovery rate (FDR; see Methods)

across all analysis centres. Local trends are evaluated at the same analysis centres shown in Fig. 1a.

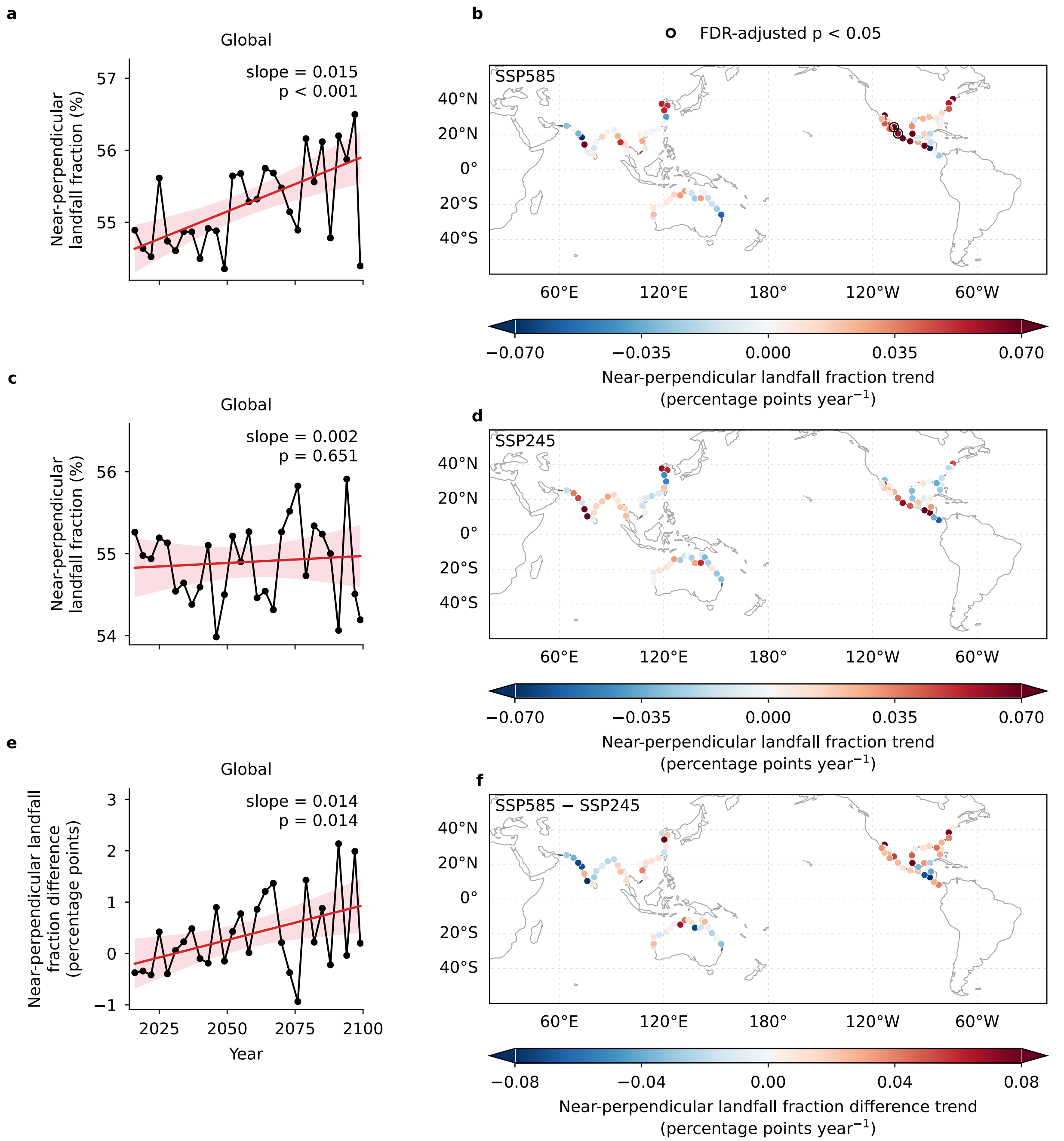

**Fig. 5 | Projected changes in the fraction of near-perpendicular tropical cyclone landfalls under future emissions scenarios. a,** Global near-perpendicular landfall (see Methods) fraction in pooled 3-year bins under SSP585 during 2015–2099**. b,** Local SSP585 trends. **c,** Global fraction in pooled 3-year bins under SSP245. **d**, Local SSP245 trends. **e**, Global difference between SSP585 and SSP245. **f**, Local trend in the SSP585-minus-SSP245 fraction difference, calculated from matched 3-year bins. In **a**, **c** and **e**, each point represents a 3-year bin over 2015–2099 (the final point represents 2099 alone), pooled across 100 PepC-Global realizations for each of the 14 CMIP6 models (Extended Data Table 1). Black lines connect the pooled fractions in **a** and **c** and

the fraction differences in **e**; red lines show weighted least-squares trends, and pink shading indicates the 95% confidence interval around each trend. In **b**, **d** and **f**, each marker shows the trend in percentage points per year, fitted to landfall fractions pooled across all models and realizations in b and d, and to the SSP585-minus-SSP245 fraction difference in f. Blue indicates a decreasing and red an increasing fraction in **b** and **d**, and a decreasing or increasing scenario difference in **f**. Black outlines mark centres at which the trend remains significant at $p < 0.05$ after controlling the false discovery rate (FDR; see Methods) across all analysis centres. Local trends are evaluated at the same analysis centres shown in Fig. 1a.

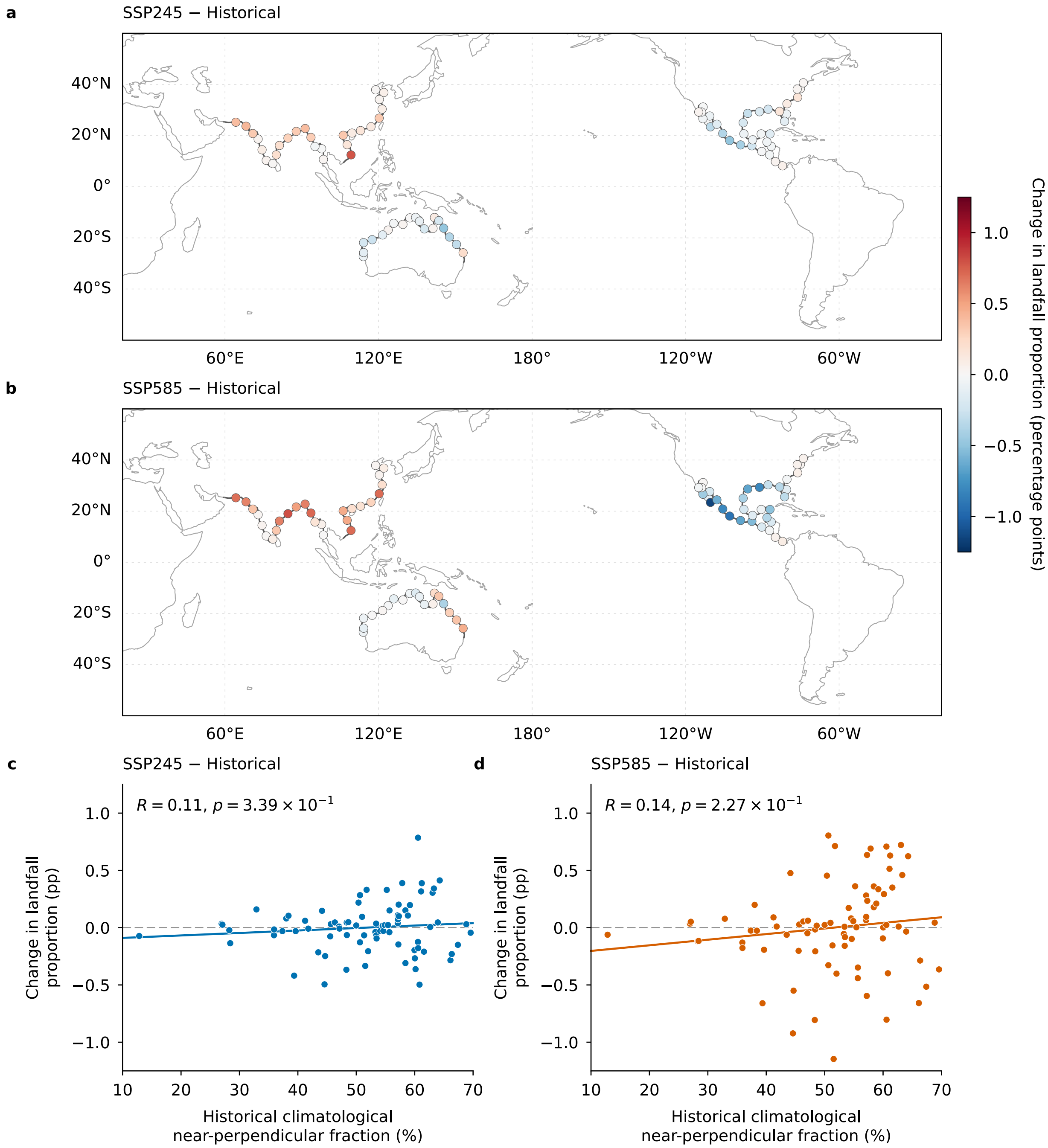

**Fig. 6 | Projected redistribution of tropical cyclone landfalls and its relationship to coastal near-perpendicular landfall fractions. a**, **b**, Changes in the proportion of global landfalls assigned to each coastal analysis centre under SSP245 in a and SSP585 in b during 2070–2099 relative to 1980–2009. Local changes are evaluated at the same analysis centres shown in Fig. 1a. Proportions are calculated for each model realization and averaged across 100 PepC-Global realizations for each of the 14 CMIP6 models. Marker colour shows the future-minus-historical change in percentage points, with blue indicating a decrease and red an increase. **c**, **d**, Relationship between the simulated historical (1980–2009) near-perpendicular landfall fraction at each centre

and the change in its share of global landfalls under SSP245 in **c** and SSP585 in **d**. Each point represents one analysis centre; solid lines show linear fits, and dashed horizontal lines mark zero change. Pearson correlation coefficients R and p-values are shown in each panel.

## METHODS

### Data

TC track data were obtained from the International Best Track Archive for Climate Stewardship (IBTrACS)[36]. Only records from US agencies (National Hurricane Center for the North Atlantic and eastern North Pacific; Joint Typhoon Warning Center for all other basins) were used. We analysed only TCs that reached a maximum sustained wind of at least 35 knots. For each of these storms, landfall detection and landfall-angle calculation used all 6-hourly positions from the first time at least 35 knots was attained through the end of the track.

Large-scale environmental fields were taken from the fifth-generation European Centre for Medium-Range Weather Forecasts (ECMWF) atmospheric reanalysis (ERA5)[40] and the NOAA National Centers for Environmental Prediction (NCEP) Global Ocean Data Assimilation System (GODAS)[41]. Monthly mean fields from both reanalyses were used to derive the predictors required by PepC-Global. Hourly ERA5 fields were additionally used to compute potential intensity, which was then aggregated to the monthly scale for model input. For future projections, monthly output from 14 CMIP6 models (Extended Data Table 1) was obtained for the historical period (1980–2014) and the SSP245 and SSP585 scenarios (2015–2099).

Large-scale predictors for the genesis, track, and intensity modules (Extended Data Fig. 6), including absolute vorticity at 850 hPa, vertical wind shear between 850 and 200 hPa, relative humidity at 600 hPa, potential intensity, zonal and meridional winds at 250 and 850 hPa, soil water content, soil temperature, and surface roughness length, were calculated from monthly ERA5 fields for the reanalysis-driven simulations and from monthly CMIP6 fields for the climate-change experiments. For ERA5, potential intensity was instead computed from hourly fields and then aggregated to monthly means, as described above. Ocean feedback factors derived from GODAS mixed-layer depth and subsurface temperature were also used. All predictors were regridded to a common 2.5° × 2.5° grid for input to PepC-Global.

## Landfall angles and near-perpendicular landfalls

For each landfall, the landfall angle is defined as the unsigned acute angle between the TC track segment crossing the coast and the local coastline segment at their intersection (Extended Data Fig. 1). Only the first intersection with the analysed continental coastline was retained. It was calculated from their WGS84 geodesic azimuths and ranges from 0° for a coast-parallel track to 90° for a perpendicular approach. Following Hall and Sobel's use of simplified coastline segments[13], we used the Natural Earth 1:110 million coastline to compute landfall angles (Extended Data Fig. 2b). Although additional observations are often available at or near landfall, we used only standard 6-hourly positions to ensure consistent comparison with the simulations; for Hurricane Sandy (2012), including the additional observations yielded a landfall angle similar to that obtained from the 6-hourly positions alone (74.27° versus 76.36°; Extended Data Fig. 2b). Following Hall and Sobel[13], we define near-perpendicular landfalls as those occurring within 30° of the coast normal, corresponding to a landfall angle of at least 60°.

## Net pre-landfall approach angles

To characterize the incoming track direction over longer intervals, we additionally calculated 6-h and 12-h net pre-landfall approach angles. For each interval, the cyclone position was interpolated to exactly 6 or 12 h before landfall, and the net approach bearing is defined as the WGS84 bearing from that position to the landfall point. The corresponding approach angle is the unsigned acute angle between this bearing and the bearing of the same coastline segment crossed at landfall (Fig. 3 and Extended Data Figs. 4 and 5). These metrics therefore represent the net displacement direction over the 6 or 12 h preceding landfall. They use the same 0–90° convention and near-perpendicular threshold as the landfall angle. Because all three angles are evaluated against the same coastline segment, their differences reflect changes in the cyclone's incoming track direction during the final hours before landfall rather than changes in the coastal reference orientation.

**PepC-Global**

PepC-Global is a probabilistic, environment-dependent framework that generates large ensembles of synthetic TCs by coupling stochastic modules for genesis, track, and intensity (Extended Data Fig. 6). The genesis module applies support vector machine classification to absolute vorticity, vertical wind shear, relative humidity and potential intensity to estimate the spatiotemporally varying genesis rate, from which individual genesis events are sampled as a Poisson process. The track module uses random forest regression with 250- and 850-hPa zonal and meridional winds, together with latitude, to predict 6-h longitude and latitude displacements, with Gaussian perturbations representing unresolved variability. Over ocean, the intensity module uses a hidden Markov model with covariates, including potential intensity, vertical wind shear, relative humidity and an ocean feedback factor, together with the current intensity and previous 6-h intensity change, to predict 6-h intensity changes. The intensity module also includes an environment-dependent over-land component, which represents intensity evolution after landfall using soil water content, soil temperature and surface roughness length. The probabilistic design of PepC-Global enables the generation of millions of physically plausible TCs. For more details of the model, refer to ref. 35.

For evaluation against observations, PepC-Global was driven by ERA5 atmospheric and GODAS oceanic fields over 1980–2024 to generate 400 stochastic realizations. For the climate-change experiments, PepC-Global was driven by monthly output from 14 CMIP6 models, using historical simulations for 1980–2014 and SSP245 and SSP585 simulations for 2015–2099. For each CMIP6 model and experiment, we generated 100 stochastic realizations.

**False-discovery-rate (FDR) control**

Testing multiple hypotheses increases the likelihood of false-positive results. We therefore applied the Benjamini–Hochberg procedure to the $m$ centres for local analysis. The p-values from all $m$ analysis centres were ranked from smallest to largest as $p_{(1)} \leq \cdots \leq p_{(m)}$. At an FDR of $q = 0.05$, the largest rank $k$ satisfying $p_{(k)} \leq \frac{k}{m} q$ was identified, and all centres with ranks from 1 to $k$ are considered significant.

## DATA AVAILABILITY

All data used in this study are publicly accessible. IBTrACS data are available from the NOAA National Centers for Environmental Information product page at https://www.ncei.noaa.gov/products/international-best-track-archive. ERA5 data are available from the Copernicus Climate Data Store, including hourly pressure-level fields https://cds.climate.copernicus.eu/datasets/reanalysis-era5-pressure-levels?tab=overview, hourly single-level fields https://cds.climate.copernicus.eu/datasets/reanalysis-era5-single-levels?tab=overview, monthly pressure-level means https://cds.climate.copernicus.eu/datasets/reanalysis-era5-pressure-levels-monthly-means?tab=overview and monthly single-level means https://cds.climate.copernicus.eu/datasets/reanalysis-era5-single-levels-monthly-means?tab=overview. GODAS data are available from the NOAA Physical Sciences Laboratory at https://psl.noaa.gov/data/gridded/data.godas.html. CMIP6 model output is available through the Earth System Grid Federation at the CMIP6 archive portals https://esgf-node.ornl.gov/search.

## CODE AVAILABILITY

The code used in this article to reproduce all figures will be made publicly available upon publication.

## ACKNOWLEDGEMENTS

We acknowledge the World Climate Research Programme, the Working Group on Coupled Modelling, and the climate modelling groups participating in CMIP6 for producing and making available the model output used in this study. This work is supported by the National Science Foundation as part of the Megalopolitan Coastal Transformation Hub (MACH) under NSF award ICER-2103754. This is MACH contribution number 115. This work is also supported by the National Oceanic and Atmospheric Administration, US Department of Commerce, under award NA23OAR4320198, at Princeton University. Cong Gao is additionally supported by a postdoctoral fellowship from the Gordon and Betty Moore Foundation.

## AUTHOR CONTRIBUTIONS

C.G. and N.L. designed the research. C.G. performed the analysis and drew all the figures. C.G. wrote the first draft of the paper. All authors provided comments on different versions of the paper.

## COMPETING INTERESTS

The authors declare no competing interests.

**Extended Data Figures 1–6 and Extended Data Table 1**

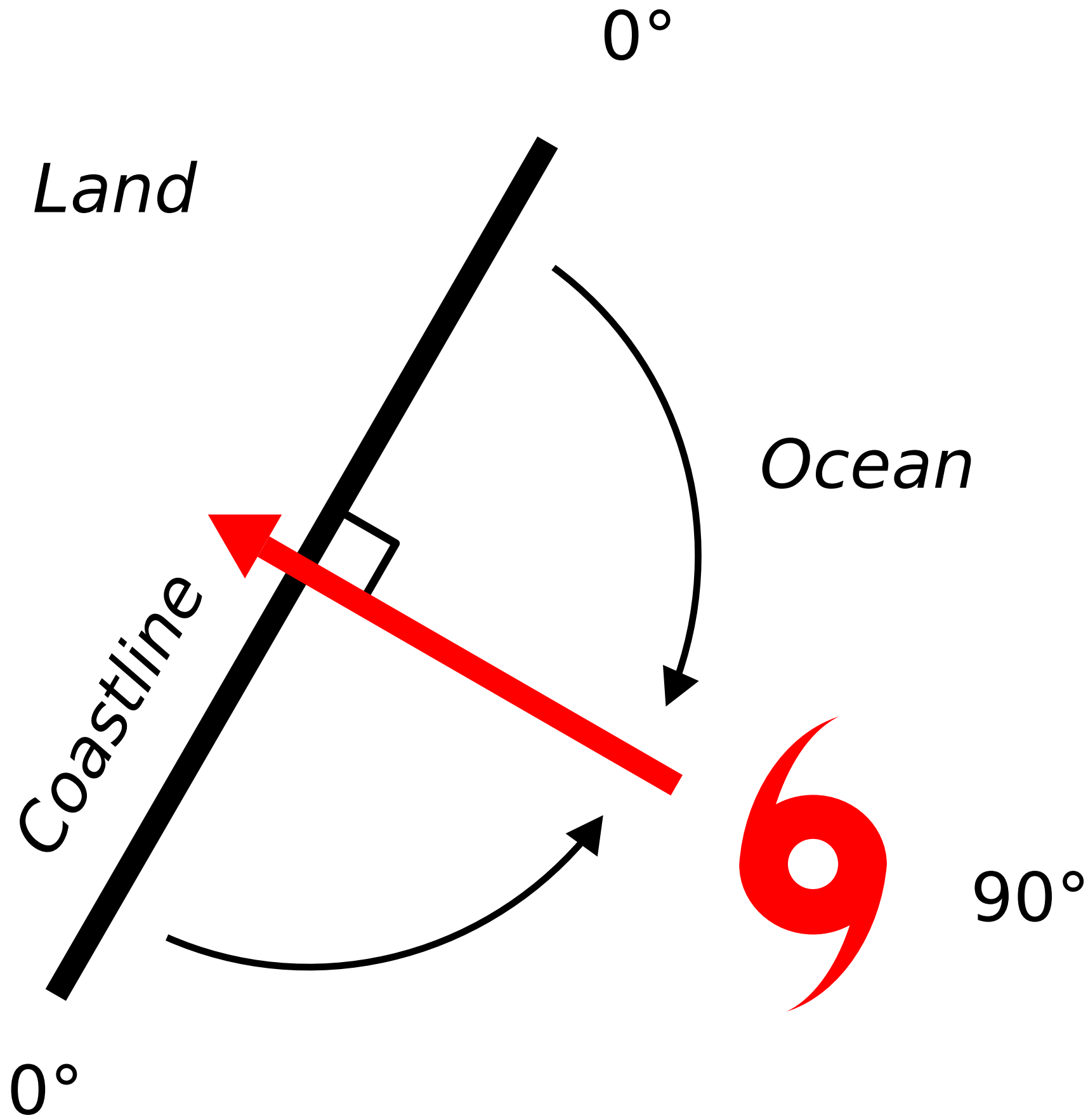

**Extended Data Fig. 1 | Definition of landfall angles of tropical cyclones (TCs).** The landfall angle is computed as the angle formed at the intersection between the coastline segment and the TC track segment defined by two adjacent positions along the storm trajectory. The crossing angle is calculated geodesically on the WGS84 reference ellipsoid, which approximates the shape of the Earth, from the azimuths of the storm-track and coastline segments at their intersection and expressed as the acute angle between them (0–90°). The schematic illustrates the coastline (thick black line) separating land from ocean, with either direction along the coastline defined as 0°. The red arrow indicates the direction of TC motion at landfall, which is always from the ocean towards the land. Curved arrows show how the landfall angle is measured from either coastline direction towards the coast-normal, ranging from 0° (parallel to the coastline) to 90° (perpendicular approach, illustrated by the red TC symbol and by the right-angle marker at the intersection of the track and the coastline).

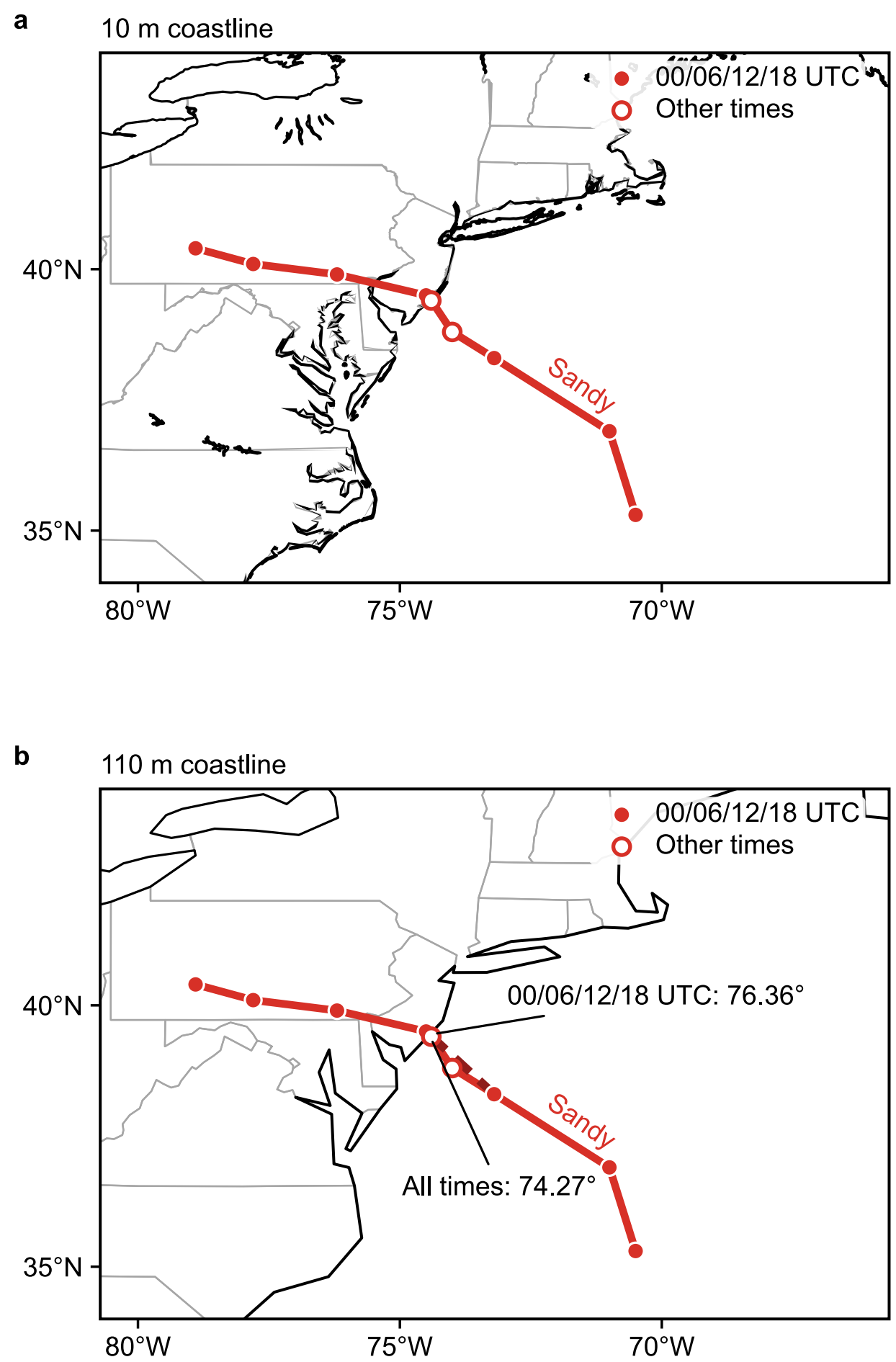

**Extended Data Fig. 2 | Example calculation of landfall angles of tropical cyclones. a**, Track of Hurricane Sandy near landfall using the Natural Earth Physical Vectors coastline at the 1:10 million scale. **b**, Corresponding calculation using the coastline at the 1:110 million scale, the coarsest and most simplified scale provided by Natural Earth. The landfall angle is calculated as defined in Extended Data Fig. 1. Red filled circles denote observations at 00, 06, 12 and 18 UTC, whereas open circles denote observations at other times. For Hurricane Sandy, additional observations were available at 21:00 and 23:30 UTC on 29 October 2012 between the 18:00 UTC observation on 29 October and the 00:00 UTC observation on 30 October, with landfall occurring between 23:30 and 00:00 UTC. The dashed line connects the 18:00 and 00:00 UTC observations. The landfall angle is 74.27° when all available observations are used and 76.36° when only observations at 00, 06, 12 and 18 UTC are used. Black lines denote coastlines.

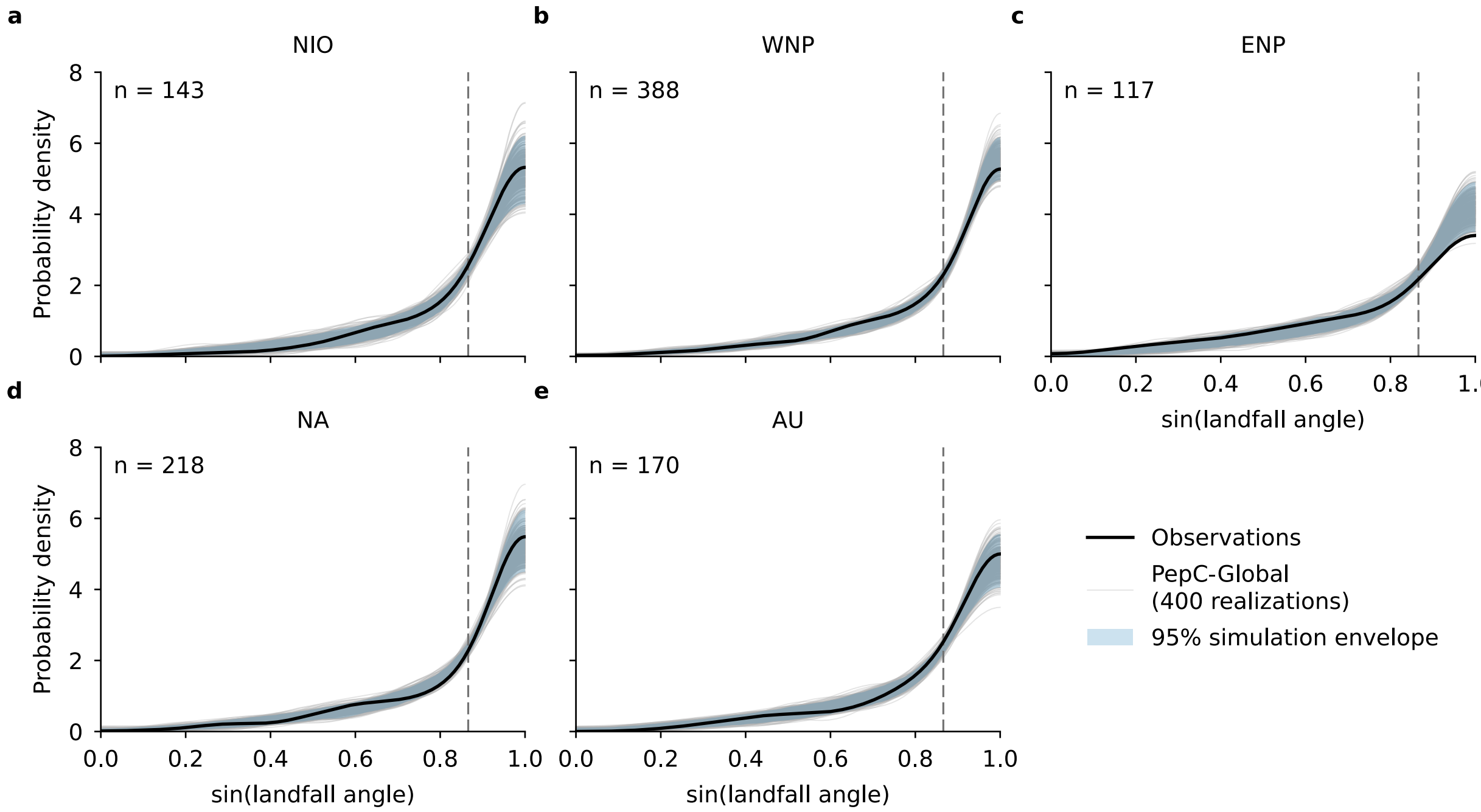

**Extended Data Fig. 3 | Observed and simulated distributions of sin of landfall angle along five continental coastlines.** **a**, North Indian Ocean (NIO). **b**, Western North Pacific (WNP). **c**, Eastern North Pacific (ENP). **d**, North Atlantic (NA). **e**, Australian Region (AU). Each panel shows the probability density of sin of landfall angle (see Methods) for tropical cyclones making landfall from 1980 to 2024. Black lines show the observed distribution, grey lines show the 400 individual PepC-Global realizations and blue shading indicates the 95% simulation envelope across those realizations; the dashed line marks the 60° near-perpendicular threshold. Here, n denotes the number of observed landfalling tropical cyclones for each coastline.

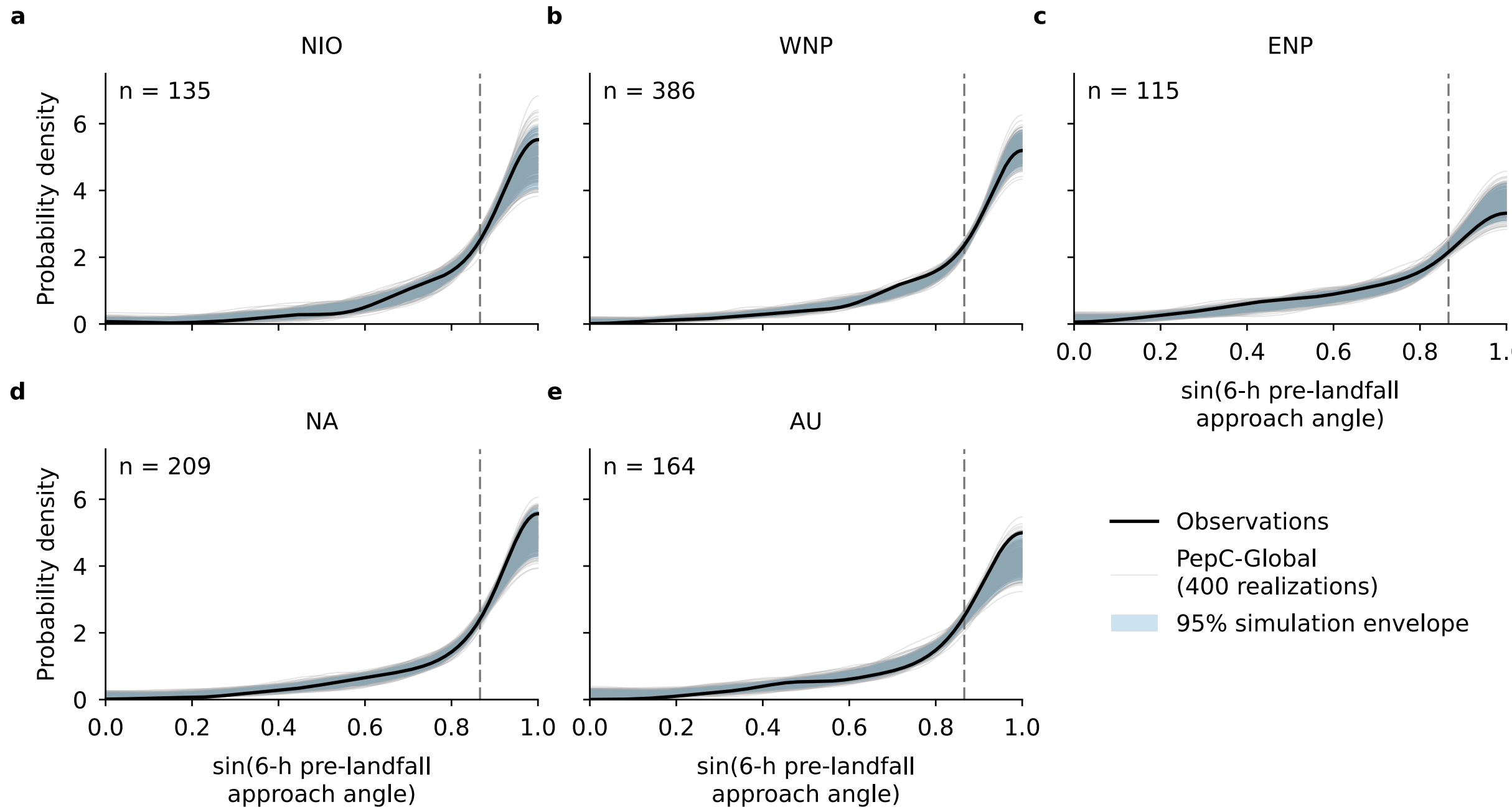

**Extended Data Fig. 4 | Observed and simulated distributions of sin of the 6-h net pre-landfall approach angle along five continental coastlines. a**, North Indian Ocean (NIO). **b**, Western North Pacific (WNP). **c**, Eastern North Pacific (ENP). **d**, North Atlantic (NA). **e**, Australian Region (AU). Each panel shows the probability density of sin of 6-h net pre-landfall approach angle (see Methods), for tropical cyclones making landfall from 1980 to 2024. Black lines show the observed distribution, grey lines show the 400 individual PepC-Global realizations and blue shading indicates the 95% simulation envelope across those realizations; the dashed line marks the 60° near-perpendicular threshold. Here, n denotes the number of observed landfalling tropical cyclones with valid 6-h net pre-landfall approach angles for each coastline.

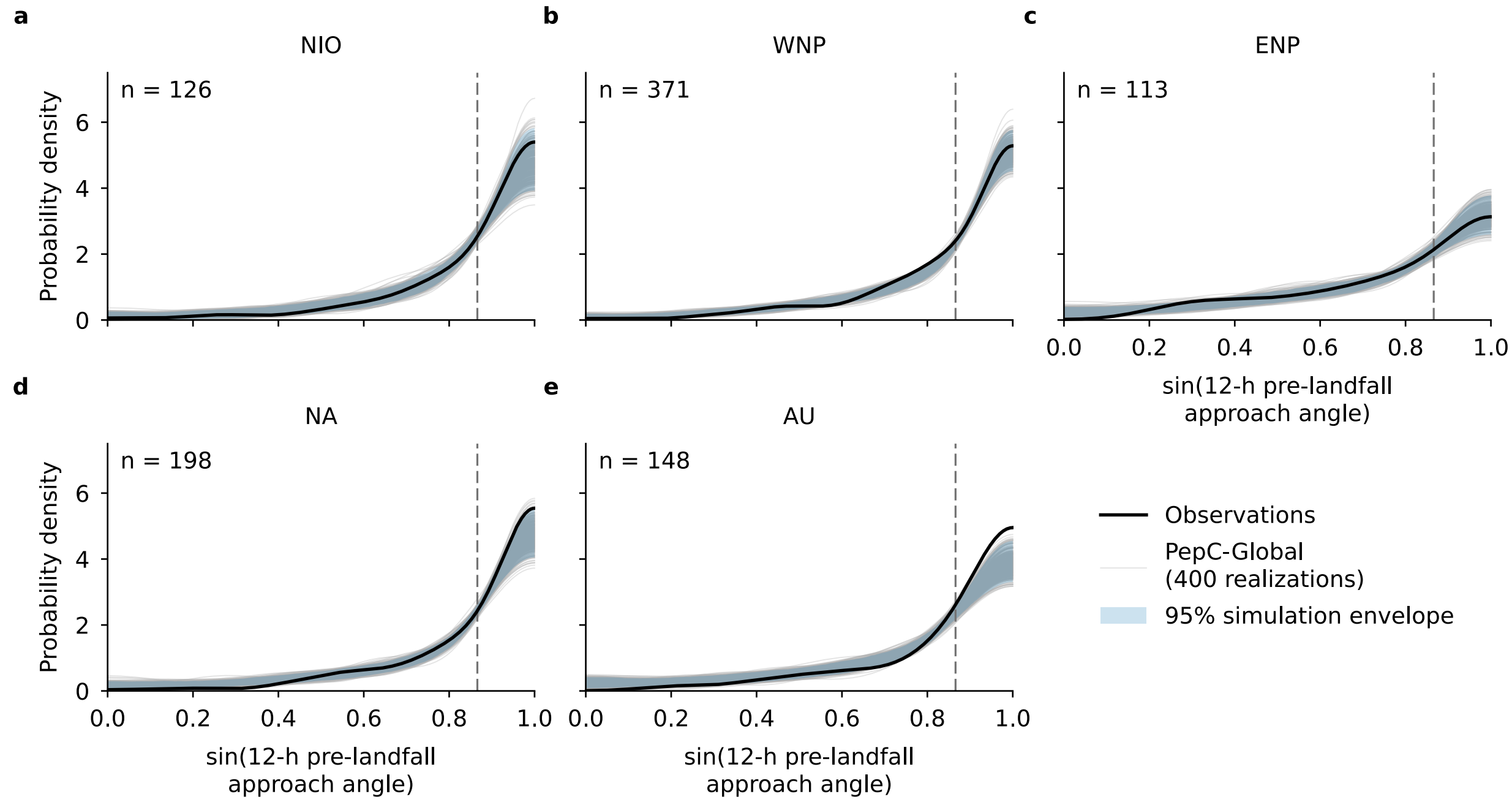

**Extended Data Fig. 5 | Observed and simulated distributions of sin of the 12-h net pre-landfall approach angle along five continental coastlines.** **a**, North Indian Ocean (NIO). **b**, Western North Pacific (WNP). **c**, Eastern North Pacific (ENP). **d**, North Atlantic (NA). **e**, Australian Region (AU). Each panel shows the probability density of sin of 12-h net pre-landfall approach angle (see Methods), for tropical cyclones making landfall from 1980 to 2024. Black lines show the observed distribution, grey lines show the 400 individual PepC-Global realizations and blue shading indicates the 95% simulation envelope across those realizations; the dashed line marks the 60° near-perpendicular threshold. Here, n denotes the number of observed landfalling tropical cyclones with valid 12-h net pre-landfall approach angles for each coastline.

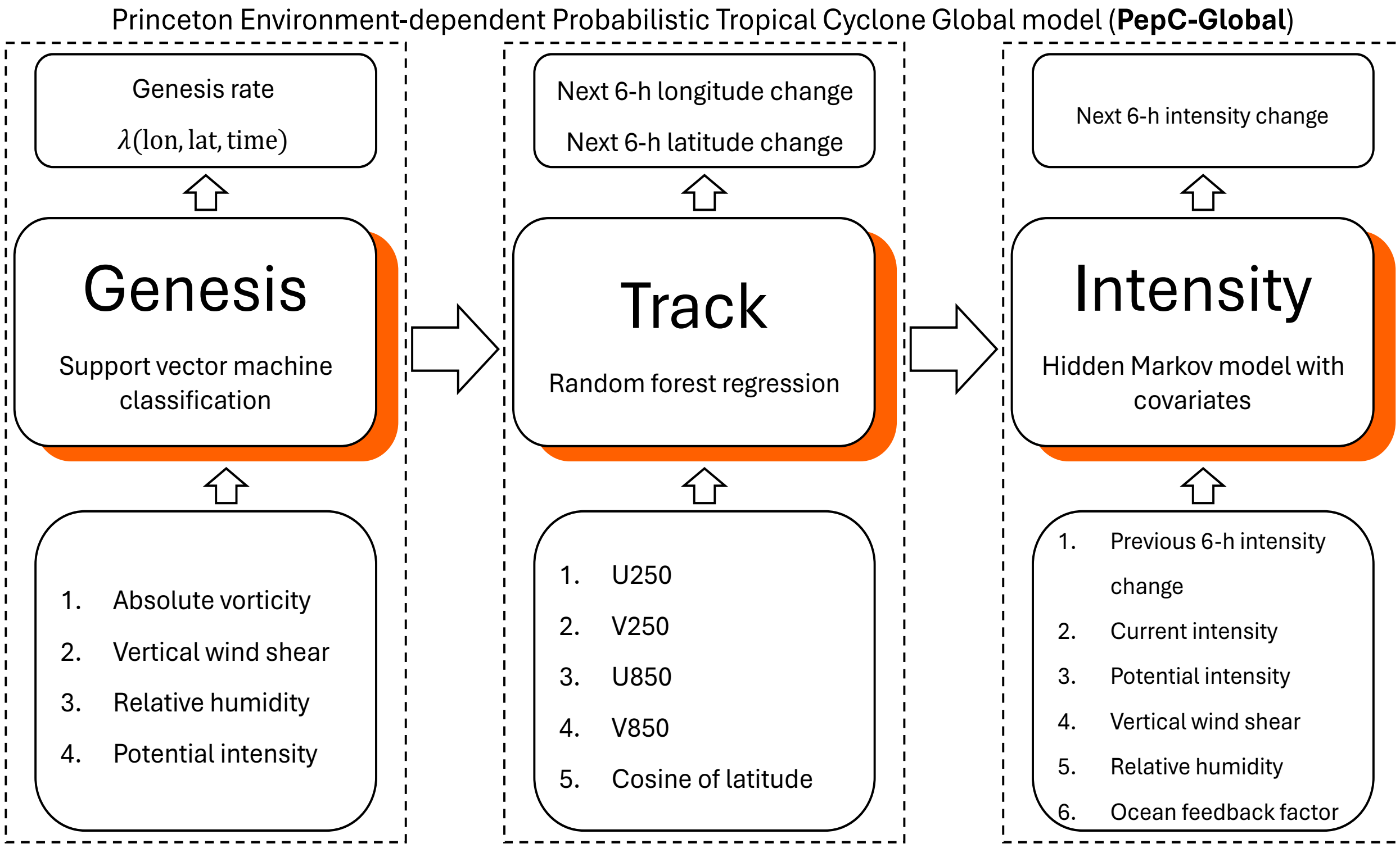

**Extended Data Fig. 6 | Schematic of the PepC-Global framework.** The Princeton Environment-Dependent Probabilistic Tropical Cyclone Global model (PepC-Global) consists of three coupled stochastic modules: genesis (support vector machine classification), track (random forest regression), and intensity (hidden Markov model with covariates). The track and intensity modules each predict the next 6-hour change in their respective variables conditioned on large-scale environmental predictors.

**Extended Data Table 1 | CMIP6 models and variants used in this study.** For each model, r1i1p1f1 is used as the default variant when available, with r1i1p1f2 selected otherwise; for CESM2, however, the SSP245 and SSP585 simulations use r10i1p1f1 because the original SSP r1i1p1f1 runs were retracted owing to a forcing-data bug and rerun with corrected forcing under new variant labels.

| Model | Historical variant | SSP245 variant | SSP585 variant |
|---|---|---|---|
| ACCESS-ESM1-5 | r1i1p1f1 | r1i1p1f1 | r1i1p1f1 |
| AWI-CM-1-1-MR | r1i1p1f1 | r1i1p1f1 | r1i1p1f1 |
| BCC-CSM2-MR | r1i1p1f1 | r1i1p1f1 | r1i1p1f1 |
| CAMS-CSM1-0 | r1i1p1f1 | r1i1p1f1 | r1i1p1f1 |
| CanESM5 | r1i1p1f1 | r1i1p1f1 | r1i1p1f1 |
| CESM2 | r1i1p1f1 | r10i1p1f1 | r10i1p1f1 |
| CNRM-CM6-1 | r1i1p1f2 | r1i1p1f2 | r1i1p1f2 |
| EC-Earth3 | r1i1p1f1 | r1i1p1f1 | r1i1p1f1 |
| FGOALS-f3-L | r1i1p1f1 | r1i1p1f1 | r1i1p1f1 |
| IPSL-CM6A-LR | r1i1p1f1 | r1i1p1f1 | r1i1p1f1 |
| MIROC6 | r1i1p1f1 | r1i1p1f1 | r1i1p1f1 |
| MPI-ESM1-2-HR | r1i1p1f1 | r1i1p1f1 | r1i1p1f1 |
| MRI-ESM2-0 | r1i1p1f1 | r1i1p1f1 | r1i1p1f1 |
| UKESM1-0-LL | r1i1p1f2 | r1i1p1f2 | r1i1p1f2 |